\documentclass{jpp}
\usepackage{graphicx}

\usepackage[utf8]{inputenc}
\usepackage[T1]{fontenc}
\usepackage{amsmath}
\usepackage{natbib}

\usepackage{amsmath,amssymb,bm}
\usepackage[dvipsnames]{xcolor}
\usepackage{dirtytalk}
\usepackage[colorlinks,linkcolor=blue,citecolor=magenta]{hyperref}
\usepackage[nameinlink]{cleveref}
\usepackage{float}

\usepackage{booktabs}
\usepackage{multirow}

\usepackage[linesnumbered,vlined,algoruled]{algorithm2e}
\crefname{algocf}{alg.}{algs.}
\Crefname{algocf}{Algorithm}{Algorithms}

\newcommand{\R}{\mathbb{R}}
\newcommand{\Bb}{\mathbf{B}}
\newcommand{\Db}{\mathbf{D}}
\newcommand{\Fb}{\mathbf{F}}

\newcommand{\nb}{\mathbf{n}}

\newcommand{\xb}{\mathbf{x}}
\newcommand{\yb}{\mathbf{y}}
\newcommand{\alphab}{\bm{\alpha}}
\newcommand{\Omegab}{\bm{\Omega}}
\newcommand{\epsilonb}{\bm{\epsilon}}
\newcommand{\lambdab}{\bm{\lambda}}

\newcommand{\Ac}{\mathcal{A}}
\newcommand{\Sc}{\mathcal{S}}

\newcommand{\nx}{{n_{\xb}}}
\newcommand{\nf}{{n_{\Fb}}}
\newcommand{\nq}{{n_{Q}}}
\newcommand{\ncm}{{n_{cm}}}
\newcommand{\nsm}{{n_{sm}}}
\newcommand{\nfp}{{n_{\text{fp}}}}
\newcommand{\nc}{{n_{C}}}

\def\code#1{\texttt{#1}}

\shorttitle{Stellarator Multi-objective Optimization}
\shortauthor{D. Bindel, M. Landreman, M. Padidar}

\title{Understanding Trade-offs in Stellarator Design with Multi-objective Optimization}

\author{David Bindel\aff{1},
  Matt Landreman\aff{2}
 \and Misha Padidar\aff{3}\corresp{\email{map454@cornell.edu}}}

\affiliation{\aff{1}Department of Computer Science, Cornell University, Ithaca, NY 14853, USA
\aff{2}Institute for Research in Electronics and Applied Physics, University of Maryland, College
Park MD 20742, USA
\aff{3}Center for Applied Mathematics, Cornell University, Ithaca, NY 14853, USA}

\begin{document}

\maketitle

\begin{abstract}
In designing stellarators, any design decision ultimately comes with a trade-off. Improvements in particle confinement, for instance, may increase the burden on engineers to build more complex coils, and the tightening of financial constraints may simplify the
design and worsen some aspects of transport. Understanding trade-offs in stellarator designs is critical in designing high performance devices that satisfy the multitude of physical, engineering, and financial criteria. In this study we show how multi-objective optimization (MOO) can be used to investigate trade-offs and develop insight into the role of design parameters. We discuss the basics of MOO, as well as practical solution methods for solving
MOO problems. We apply these methods to bring insight into the selection of two common design parameters: the aspect ratio of an ideal magnetohydrodynamic equilibrium, and the total length of the electromagnetic coils. 
\end{abstract}

\section{Problem Description}
Design criteria for stellarators stem from many physical phenomena as well as many disciplines. For instance, designs must consider fusion performance metrics from multiple models and scales, engineering specifications that determine the realizability of the design, and financial timelines which set the pace of research and construction. The collection of these criteria give rise to highly constrained design optimization problems with many, potentially competing, objectives. Any design decision ultimately comes with a trade-off. For instance, improvement in particle confinement may increase the burden on engineers to build more complex coils, and tightening of financial constraints may simplify the design and worsen some aspects of transport. When making design decisions, such as how to weight or balance objectives in an optimization or how tightly to enforce a constraint, it is critical to understand how each choice effects other aspects of the design. 

To understand trade-offs that appear in stellarator design we turn to the use of multi-objective optimization (MOO) \citep{ehrgott2005book,miettinen2012book,emmerich2018tutorial}. MOO treats problems of the form,
\begin{align}
    \min_{\xb\in\Omegab} \Fb(\xb) := (F_1(\xb), \ldots, F_{\nf}(\xb)),
    \label{eq:main_moo_problem}
\end{align}
where $\xb\in\R^{\nx}$ is a vector of design variables, $\Omegab\subseteq\R^{\nx}$ is a compact set, and $F_1,\ldots, F_{\nf}$ are, typically differentiable, objectives. MOO explores the \say{Pareto optimal} space of solutions -- those which are neither better nor worse than one another, and provide a distinct trade-off in the value of objectives -- as well as develops an understanding of how a reduction in one objective may require worsening of another.

In this study we introduce multi-objective optimization in the context of stellarator design. We discuss the basics of MOO, as well as a practical solution method for solving MOO problems: the $\epsilonb$-constraint method. We also a present continuation method which leverages a local expansion of the Pareto front to explore it efficiently. We apply these MOO methods to bring insight into the selection of two common design parameters: the aspect ratio of an ideal magnetohydrodynamic (MHD) equilibrium, and the total length of the electromagnetic coils. The aspect ratio has long been considered to have a trade-off with the degree to which a configuration is quasi-symmetric. A conjecture by \citep{GarrenBoozer2} states that quasi-symmetry can only be achieved through second order in the inverse aspect ratio, and hence that decreasing aspect ratio would result in a worsening of the degree of quasi-symmetry. Our numerical experiments suggest that this is indeed the case, but the trade-off is modest. We also explore the relationship between the total allowable length of the coil pack, and the ability for the coils to reproduce a target magnetic field. There is a natural trade-off here since longer more complex coils can reproduce more intricate fields than shorter coils. 

The paper is structured as follows: in \Cref{sec:mo_methods} we review MOO, as well as two methods for solving MOO problems. Subsequently, in \Cref{sec:numerical_experiments} we apply the optimization methods to explore two trade-off problems in stellarator design. Finally, in \Cref{sec:discussion} we look beyond our case studies to the more general use of MOO for stellarator design and discuss future directions.

We use bold characters, such as $\xb$, to denote vectors. 
We denote a vector $\xb$ with entry $j$ removed as $\overline{\xb}^j = (x_1,\ldots,x_{j-1},x_{j+1},\ldots,x_\nx)$. We compare vectors using vector inequalities: if $\xb \le \yb$ then $x_j \le y_j$ for $j=1,\ldots,\nx$. 

\section{Multi-Objective Optimization}
\label{sec:mo_methods}

Multi-objective optimization problems typically do not have a single solution, but rather an entire set of solutions. These are points which balance a trade-off between objectives: improvement in one objective implies a worsening of the other. The set of solutions to a MOO problem is called the Pareto front. In this section, we briefly formalize this notion, before introducing practical solution techniques for MOO problems.

\begin{figure}
\includegraphics[scale=0.5]{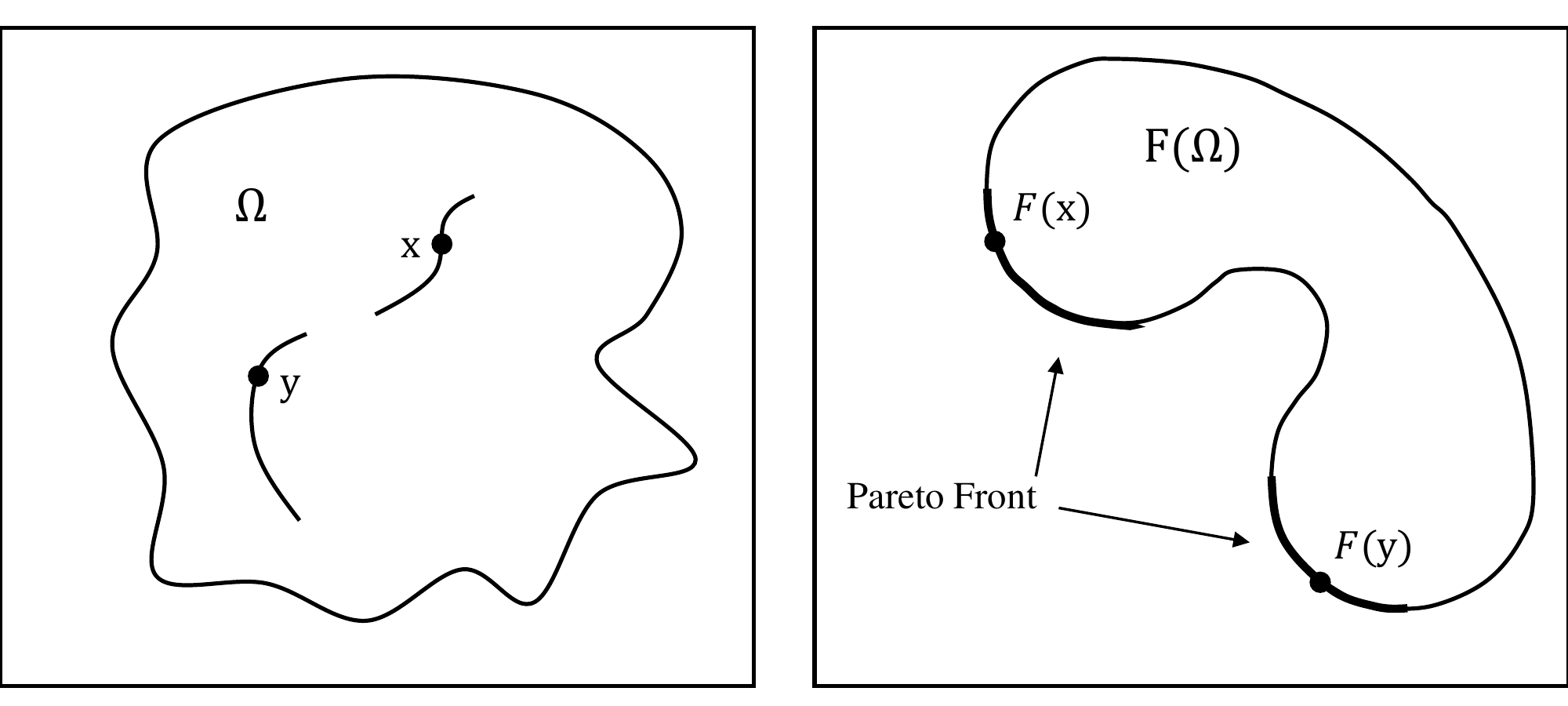}
\centering
\caption{(Left) Visualization of two points on the efficient set (black lines). (Right) Pareto front is indicated as the thicker black lines on the edge of the co-domain $\Fb(\Omegab)$.}
\label{fig:pareto_front_graphic}
\end{figure}

The idea of a trade-off can be formalized through the notion of non-dominance. A point $\xb\in\Omegab$ dominates $\yb\in\Omegab$, with respect to the objectives $\Fb$, if $\Fb(\xb) \le \Fb(\yb)$. Informally, $\xb$ is at least as good as $\yb$ in all of the objectives. If $\xb$ is not dominated by any other point in $\Omegab$, then $\xb$ is called \textit{efficient}, or Pareto Optimal, and $\Fb(\xb)$ is a \textit{non-dominated} point. The set of all efficient points forms the efficient set, and the set of all non-dominated points forms the non-dominated set, which is more commonly known as the Pareto front. Finding the Pareto front is the goal of multi-objective optimization. A visualization of an efficient set and Pareto front are shown in \Cref{fig:pareto_front_graphic}. Intuitively, efficient points are those which cannot be compared by their objective values alone. For instance, $\xb$ may have less complex coil shapes than $\yb$, but may not reproduce a target magnetic field as well. Based off of these facts alone, neither $\xb$ nor $\yb$ is a dominant configuration, and additional information would be needed to determine which configuration is preferable. 

In the absence of convex objectives, finding globally Pareto efficient solutions is difficult. Instead, we settle for searching for weakly, locally efficient points, which is all most MOO algorithms can guarantee in this setting, given a finite sample size. $\xb\in\Omegab$ is weakly efficient \citep{ehrgott2005book} if there is no $\yb\in\Omegab$ such that $\yb$ strictly dominates $\xb$, $\Fb(\yb)<\Fb(\xb)$. Moreover, $\xb\in\Omegab$ is locally efficient \citep{emmerich2018tutorial} if there exists a non-empty open ball containing $\xb$ such that $\Fb(\xb)$ is non-dominated in the intersection of the ball and $\Omegab$. We find that finding weakly, locally efficient solutions is enough to provide insight into the trade-offs of interest.

MOO problems are most commonly solved by either directly applying MOO algorithms \citep{chang2023parmoo, knowles2006parego, daulton2022multi, wagner2010expected, deb2002fast} or by reformulating the MOO problem as a series of scalar optimization problems which can be solved with scalar optimization methods. Algorithmic approaches attempt to find and explore the efficient set by taking steps to jointly minimize a combination of the objectives. These methods are a practical option for problems with many objectives, since the algorithm can weigh how to efficiently explore the high dimensional Pareto front. 
Scalarization methods, on the other hand, allow for more user involvement and incorporation of problem specific information, such as derivatives, which many MOO algorithms do not use. Scalarization methods can be formulated and solved efficiently because they rely on scalar optimization methods, and allow the user to control the exploration of the efficient set. A particularly popular scalarization approach is the \say{linearization method} \cite{ehrgott2005book}. The linearization method finds efficient points by solving the linearized problem, 
\begin{equation}
    \min_{\xb\in\Omegab} \ \ \sum_{i=1}^{\nf} \alpha_i F_i(\xb), 
    \label{eq:linearization}
\end{equation}
where the user-selected weight vector, $\alphab$, must be non-negative and sum to one. The linearization method, however, has two major drawbacks \citep{das1996closer}: 1) if the Pareto front is non-convex, then no weight vector $\alphab$ exists such that the solution to the linearized problem, \cref{eq:linearization}, lies on the non-convex portion of the Pareto front, 2) a uniformly spaced selection of weight vectors do not uniformly explore the Pareto front. For these reasons, it is worthwhile to look beyond the linearization method when solving non-convex MOO problems. 

In the remainder of this section we discuss a scalarization method, the $\epsilonb$-constraint method, which overcomes the drawbacks of the linearization method and is particularly useful for trade-offs in stellarator design. We also discuss a continuation method for locally exploring the Pareto front in an efficient manner.

\subsection{The $\epsilon$-constraint method}
\label{sec:eps_con}
Multi-objective optimization problems can be transformed into a series of scalar optimization problems via scalarization methods. One such scalarization method, the $\epsilonb$-constraint method \citep{ehrgott2005book}, finds (weakly) Pareto optimal points by minimizing a single objective subject to upper bound inequality constraints on all other objectives, with an upper bound parameter $\epsilonb\in\R^{\nf-1}$.  To find points on the Pareto front, the $\epsilonb$-constraint method solves problems of the form
\begin{align}
\label{eq:eps_con}
\begin{split} 
    \min_{\xb\in\Omegab} \ \ &F_j(\xb)
    \\
     \text{s.t.}\ \  &F_1(\xb), \ldots,F_{j-1}(\xb), F_{j+1}(\xb), \ldots F_m(\xb) \le \epsilonb,
\end{split}
\end{align}
where the user select the index $j$ and vector $\epsilonb$. The $\epsilonb$-constraint method theoretically guarantees that it can be used to find any point on the Pareto front: if $\xb$ is a (local) solution to \cref{eq:eps_con}, then it is weakly (locally) efficient. Furthermore $\xb$ is efficient if and only if it is a solution to \cref{eq:eps_con} for all choices of $j\in\{1,\ldots,\nf\}$ \citep{ehrgott2005book}. 

The $\epsilonb$-constraint method is particularly relevant to stellarator optimization since we often have have bounds on the range of the objectives, which allows us to easily select the upper bound parameter, $\epsilonb$. For example, devices often have an aspect ratio between $3$ and $10$, a coil must be at least as long as the minor circumference of the plasma, and it should not be longer than a few times that. By selecting $\epsilonb$ we can target areas of the Pareto front that we can search. This is more interpretable than using other scalarization techniques, such as a weighted sum of objectives, which requires selecting abstract weights that do not have any clear relation to the Pareto front.

\subsection{Continuation methods}
\label{sec:continuation}

\begin{figure}
\includegraphics[scale=0.5]{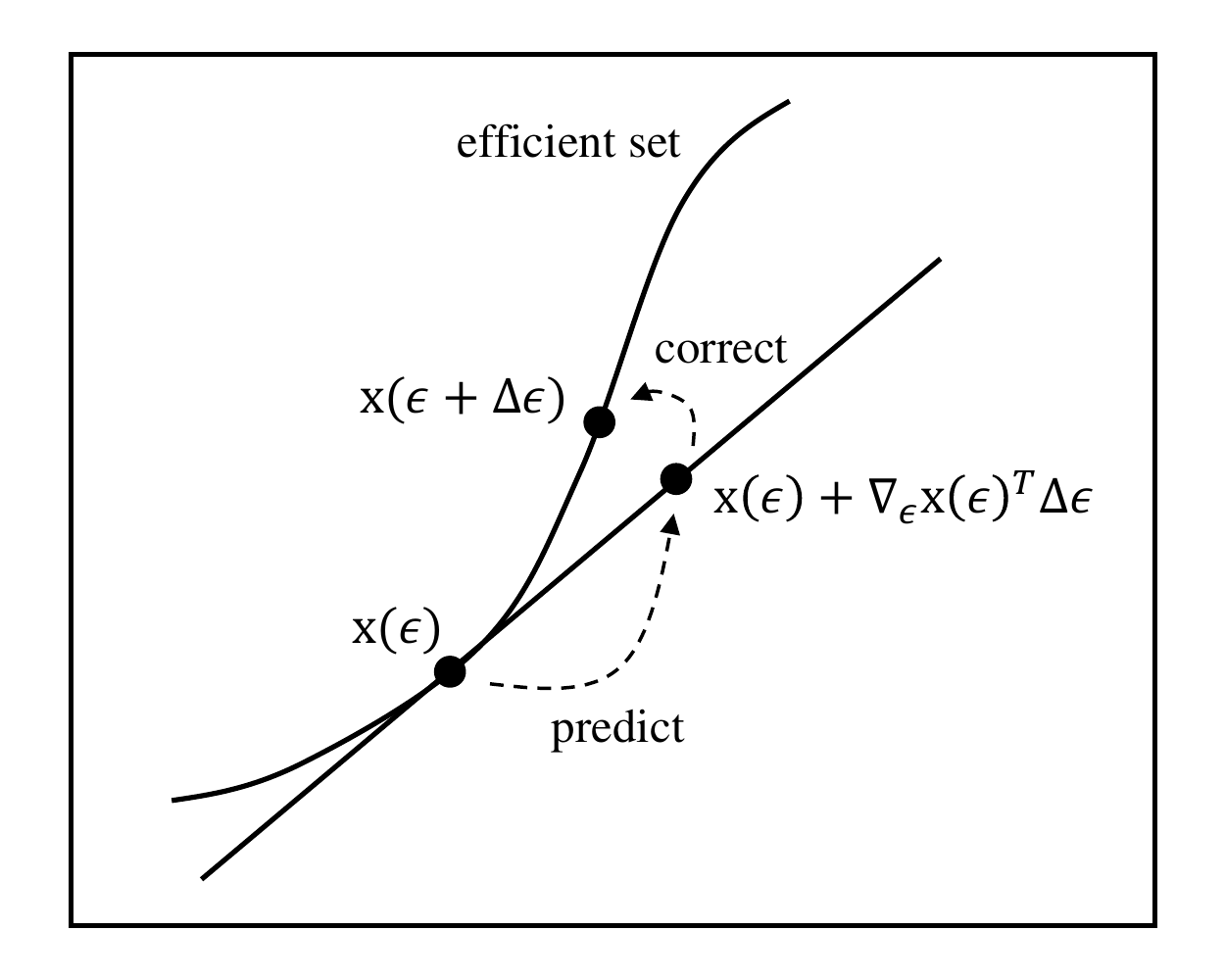}
\centering
\caption{Visualization of the predictor-corrector continuation method.}
\label{fig:continuation_method_graphic}
\end{figure}

Continuation methods \citep{hillermeier2001nonlinear,schutze2005continuation,vasilopoulos2021gradient,peitz2019multiobjective, gkaragkounis2018adjoint} in MOO explore the Pareto front locally around a given efficient point. By using optimality conditions and the Implicit Function Theorem \citep{krantz2002implicit} continuation methods build local models of the Pareto front, which they use to estimate nearby efficient points. Local exploration of the Pareto front with continuation methods comes at a relatively low computational cost when compared to restarting a scalarization solver, such as the $\epsilonb$-constraint method, from scratch. However, when combined, continuation methods and scalarization solvers make a powerful pair: the scalarization methods find efficient points over distinct parts of the Pareto front and the low cost continuation method is used to fill in the space between points. In this section, we discuss a \textit{predictor-corrector} continuation method based off of the $\epsilonb$-constraint method, which first predicts an estimate of an efficient point using a Taylor expansion, and subsequently uses the $\epsilonb$-constraint method to correct the predictions, see \Cref{alg:predictor_corrector}. While other predictor-corrector continuation schemes \citep{schutze2005continuation,vasilopoulos2021gradient,peitz2019multiobjective} write the expansion in terms of weights, or target points, we expand the Pareto front in terms of $\epsilonb$ because it is easily interpreted, and allows for the $\epsilonb$-constraint method to be used in the corrector step. 
\begin{algorithm}[tbh]
\SetAlgoNlRelativeSize{-4}
\caption{$\epsilonb$-Constraint Predictor-Corrector Method}
\label{alg:predictor_corrector}
\KwIn{Efficient point $\xb_0$, small value $\Delta \epsilonb\in\R^{\nf-1}$, index $j\in\{1,\ldots, \nf\}$.}
\KwResult{Approximate efficient points $\{\xb_k\}$.}
\For{$k=0,1,\ldots$}{
  Set $\epsilonb_{k} = \overline{\Fb}^j(\xb_k)$\;
  Compute the Lagrange multiplier $\lambda_k$ from \cref{eq:compute_lagrange} using $\epsilonb=\epsilonb_k$\;
  Compute the Hessian approximations for $\Fb$ at $\xb_k$\;
  Build the Jacobian $\Db(\xb_k,\lambda_k)$ from the Hessians approximations and gradients\;
  Solve for $\nabla_{\epsilonb}\xb$ using equation \cref{eq:kkt_deriv_matrix_system}\;
  Predict: Approximate $\xb(\epsilonb_k + \Delta \epsilonb)$ using \cref{eq:predictor}\;
  Correct: Compute $\xb_{k+1}$ with the $\epsilonb$-constraint method by solving \cref{eq:eps_con} using $\epsilonb = \epsilonb_k + \Delta \epsilonb$, and starting the optimization from the predicted point\;
  }
\end{algorithm}
The key idea of the \textit{predictor step} is to notice that the $\epsilonb$-constraint method allows us to parametrically describe weakly efficient points in terms of $\epsilonb$, i.e. we can write weakly efficient points as $\xb(\epsilonb)$. In fact, as we will show, under reasonable conditions $\xb(\epsilonb)$ is a continuously-differentiable map. The predictor step leverages this parametric representation to approximate the efficient set by a Taylor expansion. Given an efficient point $\xb$, which is a solution to \cref{eq:eps_con} with parameter $\epsilonb$, the predictor step locally approximates the efficient set by
\begin{equation}
    \xb(\epsilonb+ \Delta \epsilonb) \approx \xb(\epsilonb) +  \nabla_{\epsilonb}\xb(\epsilonb)^T\Delta \epsilonb,
    \label{eq:predictor}
\end{equation}
where $\Delta\epsilonb$ is a small change in $\epsilonb$, and $\nabla_{\epsilonb} \xb$ is the Jacobian of $\xb(\epsilonb)$ with respect to $\epsilonb$. A visualization of this method is shown in \Cref{fig:continuation_method_graphic}. In the remainder of this section we will discuss conditions which determine when this expansion exists and a method of computing the Jacobian $\nabla_{\epsilonb}\xb$. 

The existence of the expansion, \cref{eq:predictor}, and the computation of the Jacobian $\nabla_{\epsilonb}\xb$  rely on applying the Implicit Function Theorem \citep{krantz2002implicit} to the first order necessary conditions for optimality of \cref{eq:eps_con}, the Karush-Kuhn-Tucker (KKT) conditions \citep{nocedal1999numerical}. The KKT conditions are necessary conditions for a point $\xb$ to be an optimal solution to an inequality constrained optimization problem, such as the $\epsilonb$-constraint problem \cref{eq:eps_con}. In the case of the $\epsilonb$-constraint method, points satisfying the KKT conditions for \cref{eq:eps_con} are also weakly, locally efficient.

Suppose that $\xb$ is an optimal solution to \cref{eq:eps_con}, and $\Ac\subseteq \{1,\ldots,j-1,j+1,\ldots \nf\}$ is the set of $m$ active constraints, i.e. the set of constraints $i$ for which $F_i(\xb)=\epsilon_i$. 
The KKT conditions with the Linear Independence Constraint Qualification (LICQ) \citep{nocedal1999numerical} state that there exist $\lambdab\in\R^{m}$ such that $\lambdab \ge 0$ and,
\begin{align}
    \nabla F_j(\xb) + \sum_{i\in\Ac} \lambda_i\nabla F_i(\xb) &= 0,
    \label{eq:kkt_stationary}
    \\
    \lambda_i(F_i(\xb) - \epsilonb_i) &= 0 \quad \forall i\in\Ac,
    \label{eq:kkt_comp_slackness}
    \\
    F_i(\xb) &\le \epsilonb \quad \forall i\neq j.
    \label{eq:kkt_primal_feas}
\end{align}
Notice that \cref{eq:kkt_stationary,eq:kkt_comp_slackness} show that if a constraint is active and $\lambda_i >0$ then the solution of the $\epsilonb$-constraint method is dependent on $\epsilon_i$ in the sense that a small change in $\epsilon_i$ could change the solution to the system of equations. In this case, it is reasonable to write $\xb$ as a function of $\epsilonb_{\Ac} = (\epsilon_i)_{i\in\Ac}$ to stress the dependence of solutions on the parameters, $\xb(\epsilonb_{\Ac})$. The Implicit Function Theorem gives us the mathematical leeway to do so.

Informally, the Implicit Function Theorem states that if there exists a point $\xb^*$, with Lagrange multipliers $\lambdab^*$ and constraint parameters $\epsilonb_{\Ac}^*$, which satisfy the KKT conditions, and if the Jacobian, $\Db(\xb^*,\lambdab^*)$,
%
%
of the left-hand side of \cref{eq:kkt_stationary,eq:kkt_comp_slackness} with respect to $(\xb,\lambdab)$ is invertible, then $\xb(\epsilonb_{\Ac}),\ \lambdab(\epsilonb_{\Ac})$ are continuously differentiable functions of $\epsilonb_{\Ac}$ around $\epsilonb_{\Ac}^*$. Taking the form of a continuously differentiable function, the space of solutions to the $\epsilonb$-constraint problem \cref{eq:eps_con} can be approximated with the first order Taylor expansion, \cref{eq:predictor}. The Jacobian $\nabla_{\epsilonb}\xb$ can be computed by solving the system,
\begin{equation}
    \Db(\xb^*,\lambdab^*)
    \begin{bmatrix}
    \nabla_{\epsilonb}\xb \\ \nabla_{\epsilonb}\lambdab
    \end{bmatrix}
    =
    \begin{bmatrix}
        \mathbf{0} \\ \lambdab^*
    \end{bmatrix}
    \label{eq:kkt_deriv_matrix_system}
\end{equation}
The system \cref{eq:kkt_deriv_matrix_system} is derived by treating $\xb$ and $\lambdab$ as a function of $\epsilonb$ and differentiating \cref{eq:kkt_stationary,eq:kkt_comp_slackness} with respect to $\epsilonb$. Importantly, $\Db$ depends on second derivatives of the objective functions. If $\Db$ is invertible, then then the Inverse Function Theorem guarantees existence of the expansion \cref{eq:predictor}, and a unique solution exists to \cref{eq:kkt_deriv_matrix_system} \citep{krantz2002implicit}. 

Invertibility of the Jacobian matrix $\Db$ occurs naturally when $\xb,\lambdab$ satisfy the KKT conditions, and $\xb$ is a strict local minimum: $\nabla^2 F_j(\xb) + \sum_{i\neq j}\lambda_i \nabla^2 F_i(\xb)$ is positive definite over the orthogonal subspace to $\{\nabla F_i\}$ for all $i\in\Ac$ such that $\lambda_i>0$. On the other hand, the expansion does not exist if there are no active constraints: if all $F_i(\xb) < \epsilon_i$ then a small variation in any $\epsilonb_i$ will not effect the solution. Furthermore, if some constraint $i$ is active but the associated Lagrange multiplier $\lambda_i =0$, then a small increase in $\epsilon_i$ will not change the solution to \cref{eq:eps_con}, but a small decrease would. In this case, $\lambda_i$ is only sub-differentiable \citep{clarke1990optimization} with respect to $\epsilon_i$. While an expansion can be derived in this case, we ignore it for simplicity, as this case only appears at the edges of the Pareto front. We are primarily concerned with the case when all active constraints have stictly positive Lagrange multipliers, since in this case there is a trade-off to be made between objectives: reduction in $F_j$ comes at the cost of increase in $\epsilon_i$.

Now that we have established when a local expansion of the efficient set exists, and how to compute it, we can discuss the predictor-corrector method. The predictor step of the predictor-corrector method takes as input an efficient point $\xb$ and a step size $\Delta \epsilonb$. Subsequently, we compute the Lagrange multiplier $\lambdab$. If some components of $\lambdab$ are strictly positive, then we compute $\Db(\xb,\lambdab)$, solve \cref{eq:kkt_deriv_matrix_system} for $\nabla_{\epsilonb}\xb$, then predict a weakly efficient point via the expansion, \cref{eq:predictor}.

For small enough $\Delta \epsilonb$, the prediction will typically be a very good estimate of an efficient point. Nonetheless, it is helpful to finely resolve the point, particularly as this allows us to compute the subsequent predictor step accurately. The \textit{corrector step} corrects the prediction by simply solving the $\epsilonb$-constraint problem, \cref{eq:eps_con}, with new constraint parameters $\epsilonb + \Delta \epsilonb$. This method is detailed in \Cref{alg:predictor_corrector}.

\subsubsection{Computation of the Lagrange Multiplier}
In practice, we rarely have a point $\xb$ that exactly satisfies any constraint with equality. Typically, an inequality constraint may approximately be satisfied with equality, i.e. $\overline{F}^j(\xb) = \epsilonb - \nu$ for a small value of $\nu > 0$. In this case, theoretically the constraint is not active and so the Lagrange multipliers are equal to zero. However, the numerics suggest that the constraint is active and the Lagrange multiplier should not be zero. In practice, we determine the active set $\Ac$ as the set of all constraints with $\nu_i< \tau$ for some small tolerance $\tau$. The Lagrange multiplier can then be computed by solving the KKT conditions for a perturbed version of \cref{eq:eps_con} where the $\epsilon_i$ is shifted to $\epsilon_i - \nu$ for all constraints in $\Ac$. In this way, the constraints hold with exact equality, since $\overline{F}^j(\xb) = \epsilonb - \nu$, and we are justified in computing the potentially non-zero Lagrange multipliers. The Lagrange multipliers are then estimated as the solution to 
\begin{equation}
    \min_{\lambdab\geq 0} \|\nabla F_j(\xb) + \sum_{i\in\Ac}\lambda_i\nabla F_i(\xb)\|^2.
    \label{eq:compute_lagrange}
\end{equation}
Importantly, this procedure affects the prediction step in that the expansion is taken around $\epsilonb + \nu$. However, this is accounted for in step 1 of \Cref{alg:predictor_corrector}.

\section{Numerical Experiments}
\label{sec:numerical_experiments}

In this section we use MOO to bring insight into the selection of two common stellarator design parameters:
the aspect ratio of an ideal MHD equilibrium, and the total
length of the electromagnetic coils. 

Our first experiment attempts to determine if there is a trade-off between achieving precise quasi-symmetry in a \say{stage-one} stellarator design (an ideal MHD equilibrium), and having low aspect ratio. The Garren-Boozer conjecture \citep{GarrenBoozer2} suggests that exact quasi-symmetry is only possible at high aspect ratio. In addition, precise quasi-symmetry has been achieved throughout a volume in high aspect ratio stellarators \citep{landreman2022magnetic, wechsung2022precise, giuliani2022direct, landreman2022mapping}. However, it is unclear at what rate quasi-symmetry decays as the aspect ratio is increased, or if this trade-off applies to precise quasi-symmetry at all. Our first experiment answers the following question: \say{To what extent does the aspect ratio limit the degree to which quasi-symmetry can be achieved throughout a volume?}

Our second experiment considers a trade-off in the \say{stage-two} design problem, where coils are optimized to fit a target magnetic field. A target magnetic field can be recreated arbitrarily well by coils which have no constraint on their length. However, when restricted to have a reasonably short length for engineering purposes, coils may not be able to reproduce a target magnetic field. In this problem, we aim to understand how reduction in the allowable coil length worsens the reproduction of the target magnetic field.

\subsection{Problem 1: the aspect ratio and quasi-symmetry trade-off}
\label{sec:aspect_quasisymmetry_experiment}

We seek a plasma boundary shape, parameterized by $\xb$, of an Ideal MHD equilibrium for a quasi-helical (QH) stellarator configuration that has minimal aspect ratio $A(\xb)$ and deviation from quasi-symmetry $Q_{M,N}(\xb)$,
\begin{equation}
    \min_{A(\xb)\in[A_l,A_u]} (A(\xb),Q_{M,N}(\xb)).
    \label{eq:aspect_qs_prob}
\end{equation}
Including bound constraints on the aspect ratio $A_l=3, A_u=10$, restricts our decision space to a realistic range of configurations. For convenience, we collect the $\nf=2$ objectives into the vector $\Fb(\xb) :=(A(\xb),Q_{M,N}(\xb))$.

\begin{figure}
\includegraphics[scale=0.4]{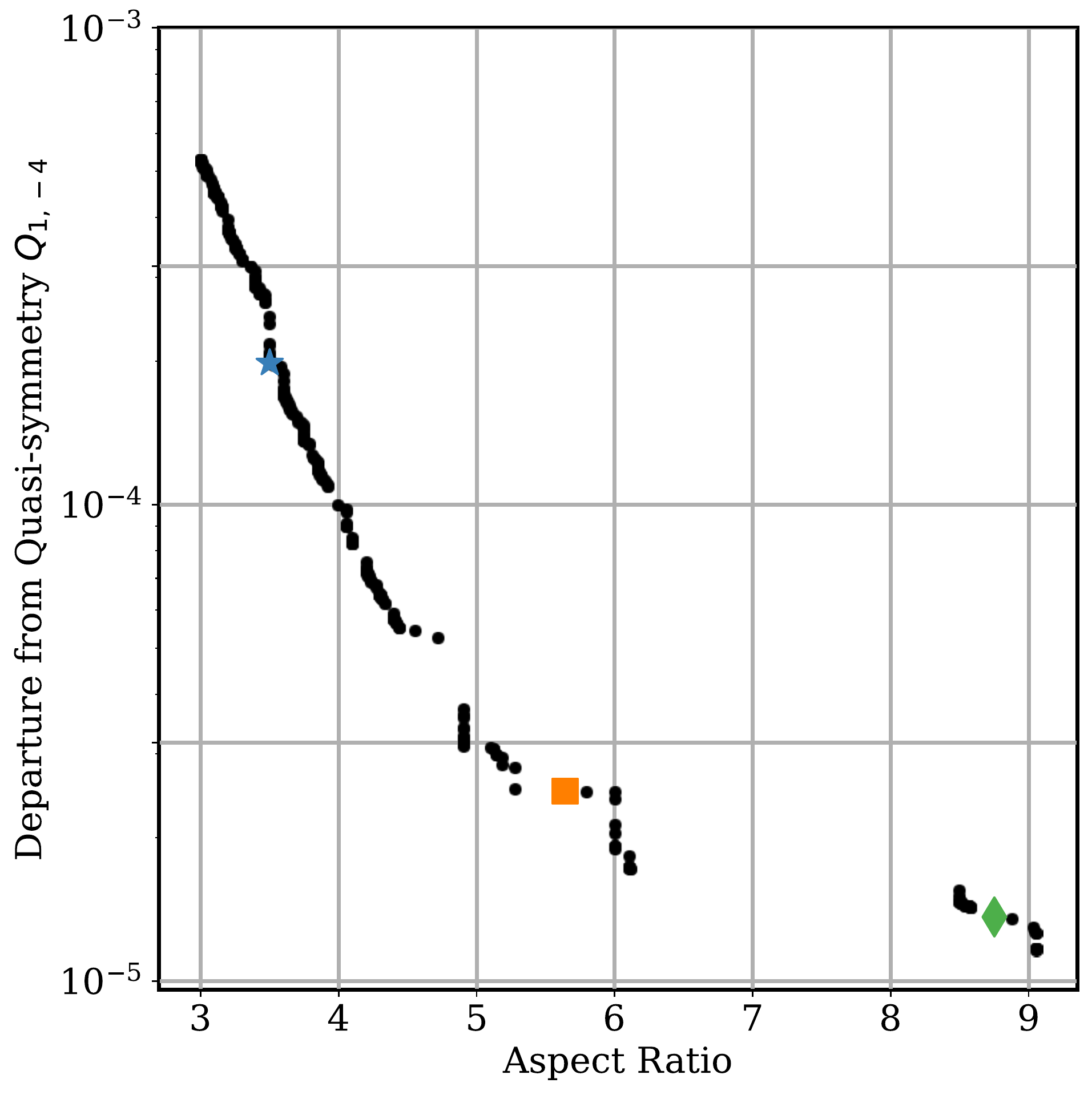}
\centering
\caption{Pareto front of the aspect ratio and quasi-symmetry objectives over the domain $A_l \le A(\xb) \le A_u$. The configurations corresponding to the blue star, orange square and green diamond markers are plotted in \Cref{fig:3d_plots}. We find that the efficient set undergoes a branch change as the aspect ratio crosses the large gap from $\approx 6.11$ to $\approx 8.5$.}
\label{fig:aspect_qs_pareto_front}
\end{figure}

The violation of quasi-symmetry is defined as the \textit{Quasi-symmetry Ratio Residual} objective employed in \citep{landreman2022magnetic}. The objective measures the departure from quasi-symmetry throughout the plasma volume as the sum of flux surface averages across surfaces $(s_1,\ldots,s_{n_s})$, 
\begin{equation}
    Q_{M,N}(\xb) = \sum_{s_j} \left\langle \left( \frac{1}{B^3}[(N-\iota M)\Bb\times\nabla B\cdot\nabla\psi - (MG-NI)\Bb\cdot\nabla B]\right)^2\right\rangle.
    \label{eq:quasisymmetry_objective}
\end{equation}
The helicity parmeters, $M,N$, determine the type of quasi-symmetry which $Q_{M,N}$ measures: $M=1,N=0$ being quasi-axisymmetry, and $M=1, N=\pm k\nfp$ with non-zero integers $k$ being quasi-helical symmetry \citep{landreman2022magnetic}. For our experiments, the helicity parmeters were set to $M=1,N=-\nfp$ for the $\nfp=4$ field period magnetic field. Discretization of the flux surface average over the poloidal and toroidal angles results in an objective with sum-of-squares structure, $Q_{M,N}(\xb) = \sum_{i=1}^{\nq} Q_{M,N,i}^2(\xb)$, where $Q_{M,N,i}$ measures the violation of quasi-symmetry at a point throughout the volume. For a given plasma boundary, $\xb$, we compute the aspect ratio using the definition in the Variational Moments Equilibrium code (\code{VMEC}) \citep{hirshman1986three,hirshman1983steepest},
\begin{equation}
    A(\xb) = R_{\text{major}}(\xb)/R_{\text{minor}}(\xb) \quad R_{\text{minor}} = \sqrt{\overline{C}_A(\xb)/\pi} \quad R_{\text{major}(\xb)} = \frac{V(\xb)}{2\pi^2R_{\text{minor}}^2(\xb)}
\end{equation}
where $\overline{C}_A$ is the average cross-sectional area of the surface and $V$ is the volume enclosed by the surface.

The decision variables, $\xb$, are Fourier amplitudes that describe the shape of the plasma boundary. The plasma boundary is represented in terms of the standard cylindrical coordinates $(R,\phi,Z)$ where $R,Z$ are parameterized as a Fourier series in the poloidal and toroidal angles $\theta$ and $\phi$,
\begin{equation}
\begin{split} 
    R(\theta,\phi) &= \sum_{n=0}^{\nsm} R_{0,n}\cos(- n_{\text{fp}}n\phi) + \sum_{m =1}^{\nsm} \sum_{n=-\nsm}^{\nsm} R_{m,n}\cos(m\theta - n_{\text{fp}}n\phi),
    \\
    Z(\theta,\phi) &= \sum_{n=1}^{\nsm} Z_{0,n}\sin(- n_{\text{fp}}n\phi)  + \sum_{m =1}^{\nsm} \sum_{n=-\nsm}^{\nsm} Z_{m,n}\sin(m\theta - n_{\text{fp}}n\phi).
\end{split}
    \label{eq:vmec_fourier_rep}
\end{equation}
The number of modes describing the surface $\nsm = \{0,1,2,\ldots\}$ can be increased to achieve more intricate boundary representations. Field period symmetry with $\nfp$ periods and stellarator symmetry have been assumed.  
The major radius $R_{0,0}$ is held fixed throughout the optimization, to fix scale of the design. The Fourier amplitudes are collected into the decision variable via $\xb = (R_{0,1},\ldots, Z_{0,0},\ldots)$. The total number of decision variables satisfies $\nx = 4\nsm^2 + 4\nsm$. 

Numerical experiments were performed using \code{SIMSOPT} \citep{landreman2021simsopt} to handle variables, compute objectives, and interface with \code{VMEC} which computed the Ideal MHD equilibria from the plasma boundary representation. 
The $\epsilonb$-constraint method was used to find points along the Pareto front where quasi-symmetry was set to be the target function for minimization while the aspect ratio was constrained by $\epsilon$. The bound constraints on the aspect ratio made specifying $\epsilon$ straight forward: $\epsilon$ was set to linearly spaced values between $A_l$ and $A_u$. The $\epsilonb$-constraint problem \cref{eq:eps_con} was solved by reformulating the constrained optimization problem with a quadratic penalty method \citep{nocedal1999numerical}. The penalty parameter was increased from an initial value of $1$ by a factor of $10$ at each iteration, and the quadratic penalty subproblems were solved by applying a Gauss-Newton optimization routine. \code{SIMSOPT} was used to compute forward difference gradients via MPI-based concurrent function evaluations. To avoid local minima, the number of Fourer modes, $\nsm$, was increased iteratively from $1$ to $5$, reaching $\nx=120$ variables. The $\epsilon$-constraint problem was solved with the penalty method after each increase of $\nsm$.


\begin{figure}
\includegraphics[scale=0.35]{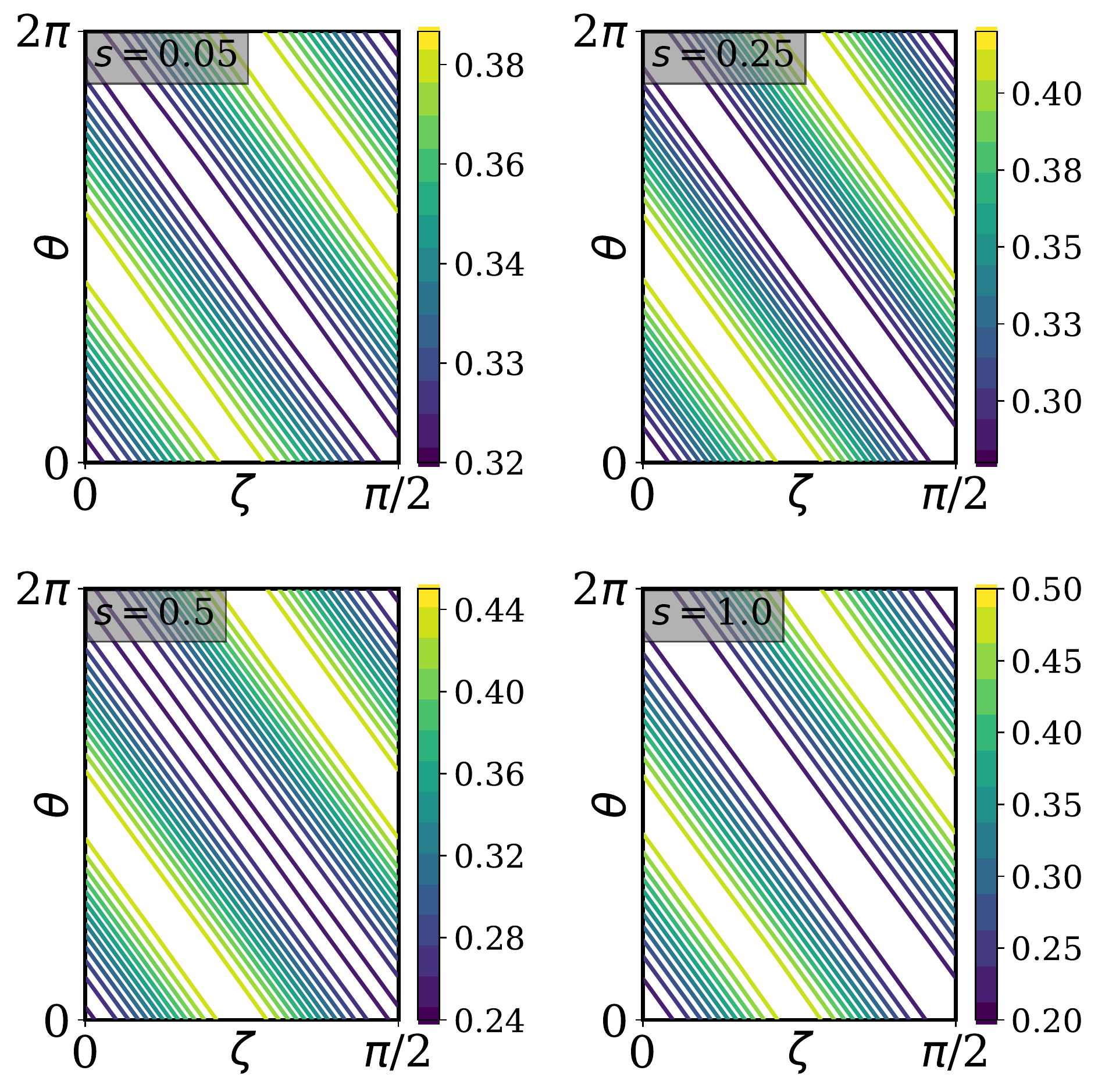}
\includegraphics[scale=0.3]{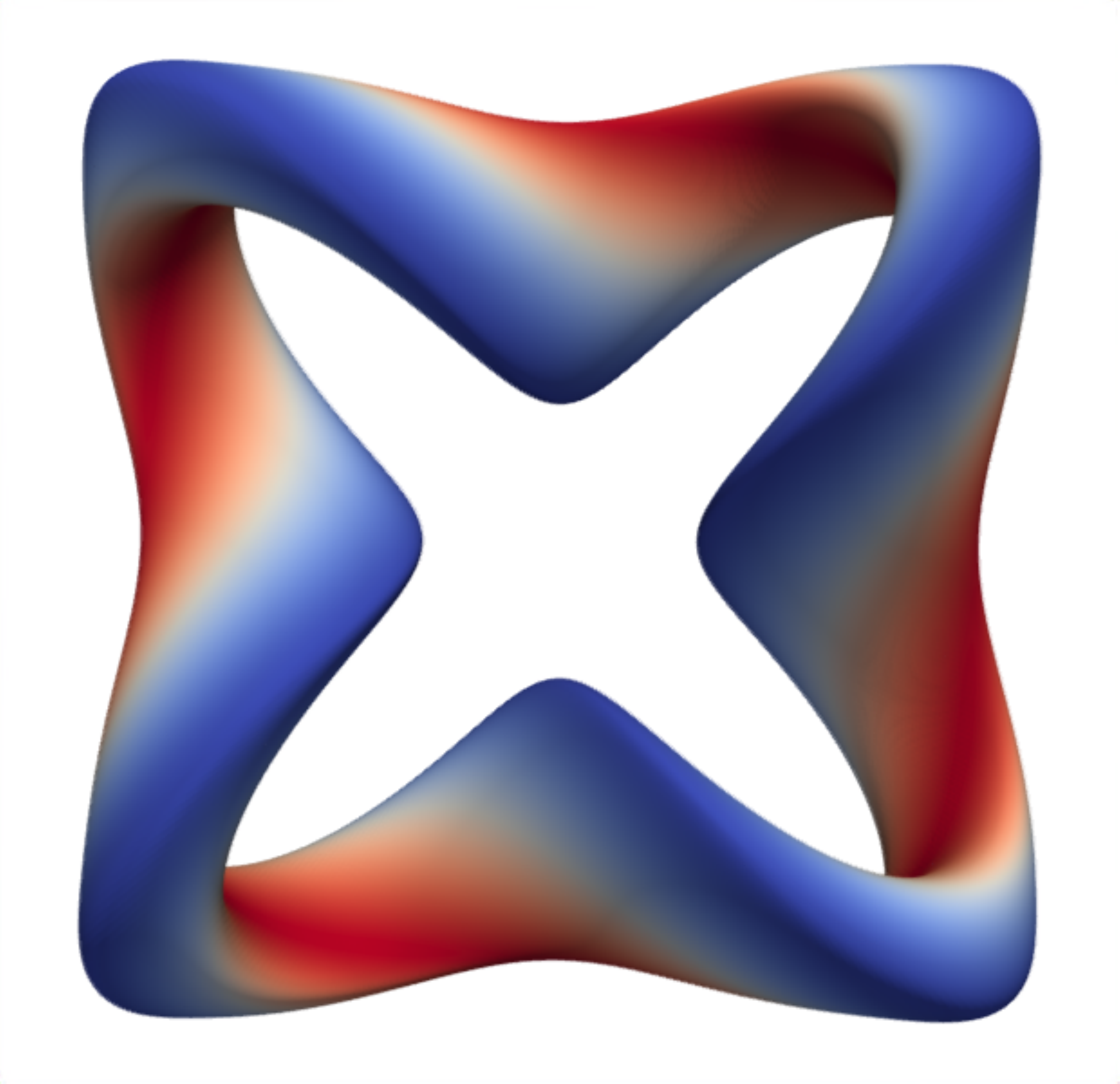}
\includegraphics[scale=0.35]{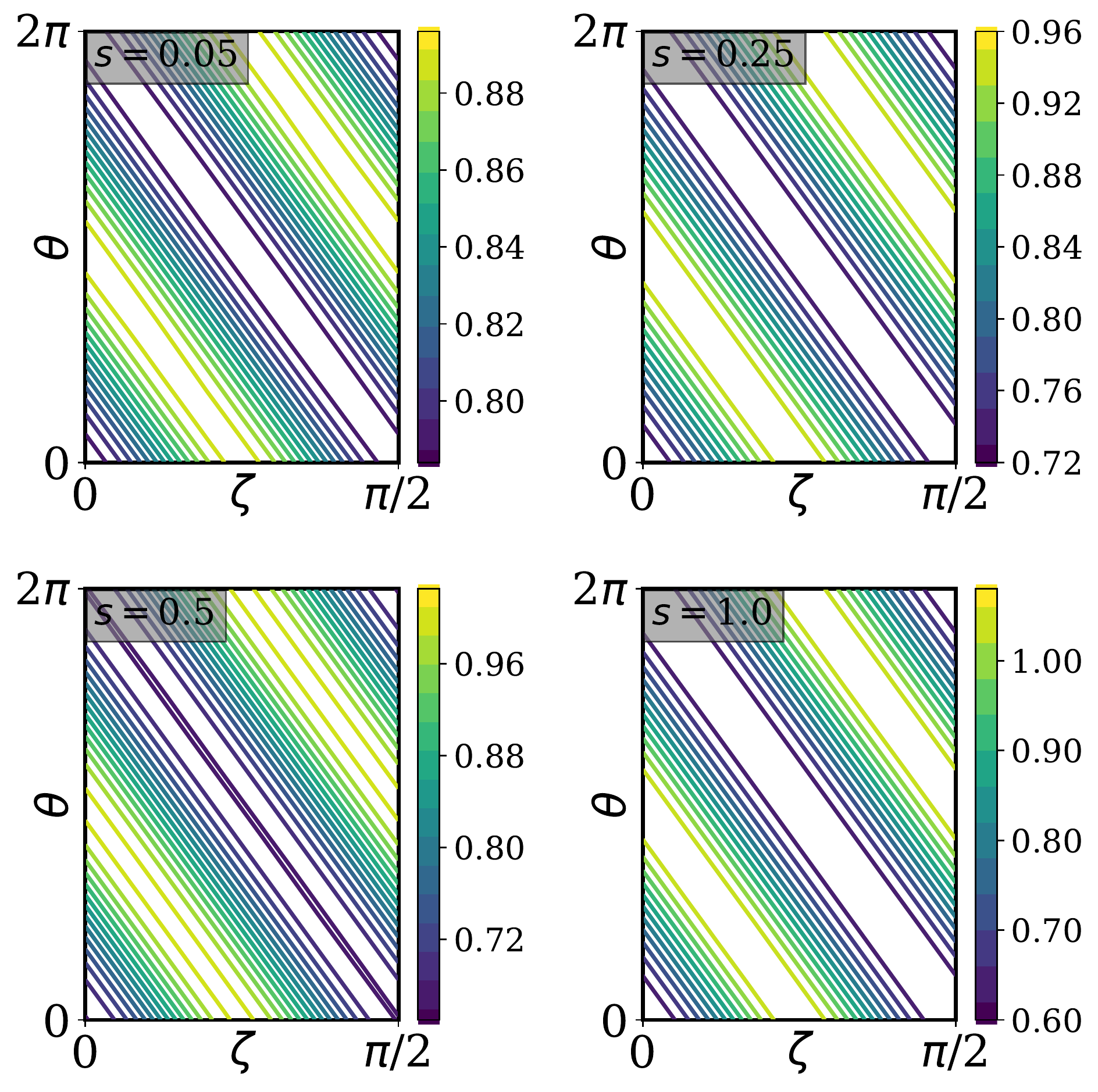}
\includegraphics[scale=0.3]{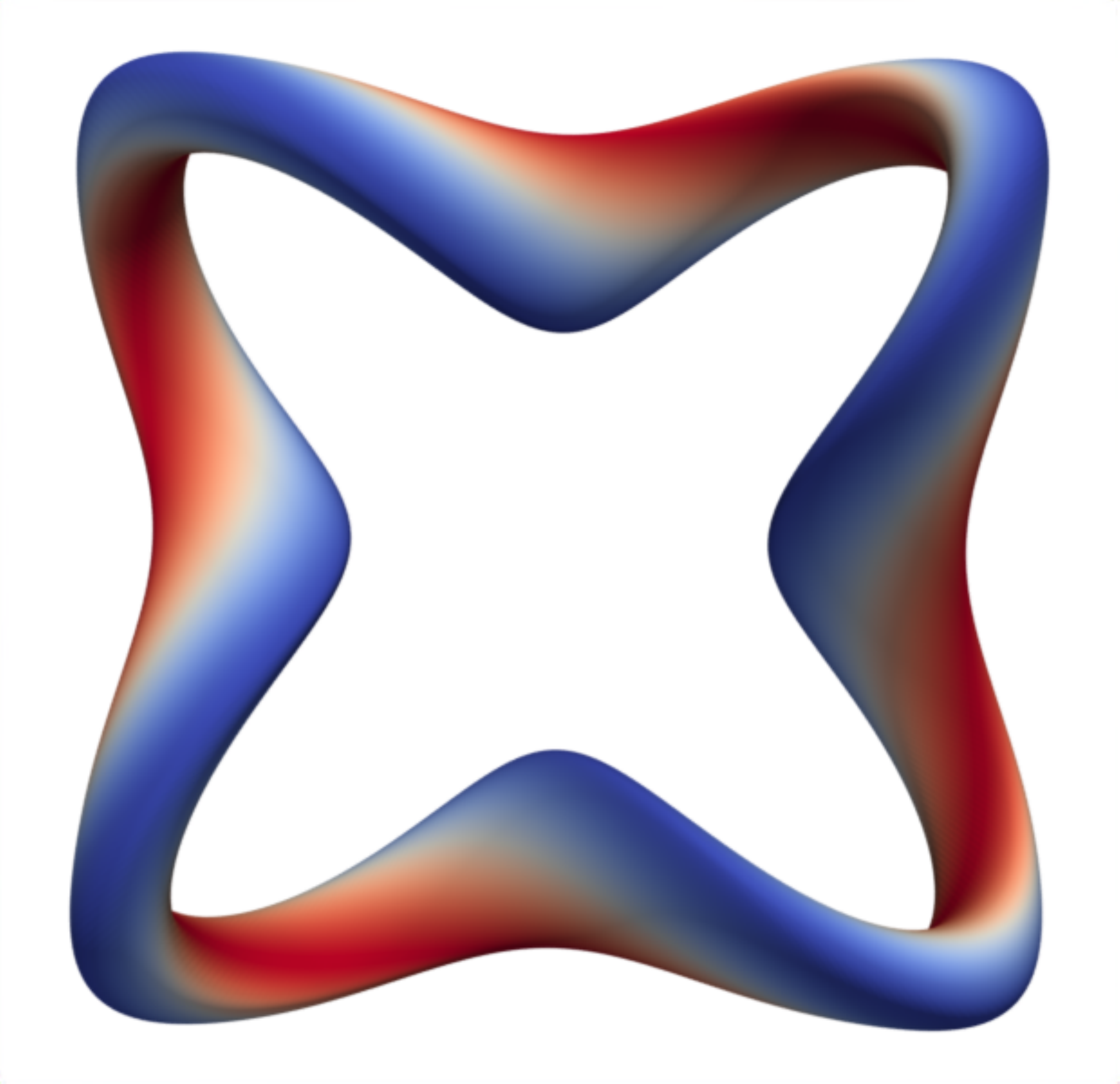}
\includegraphics[scale=0.35]{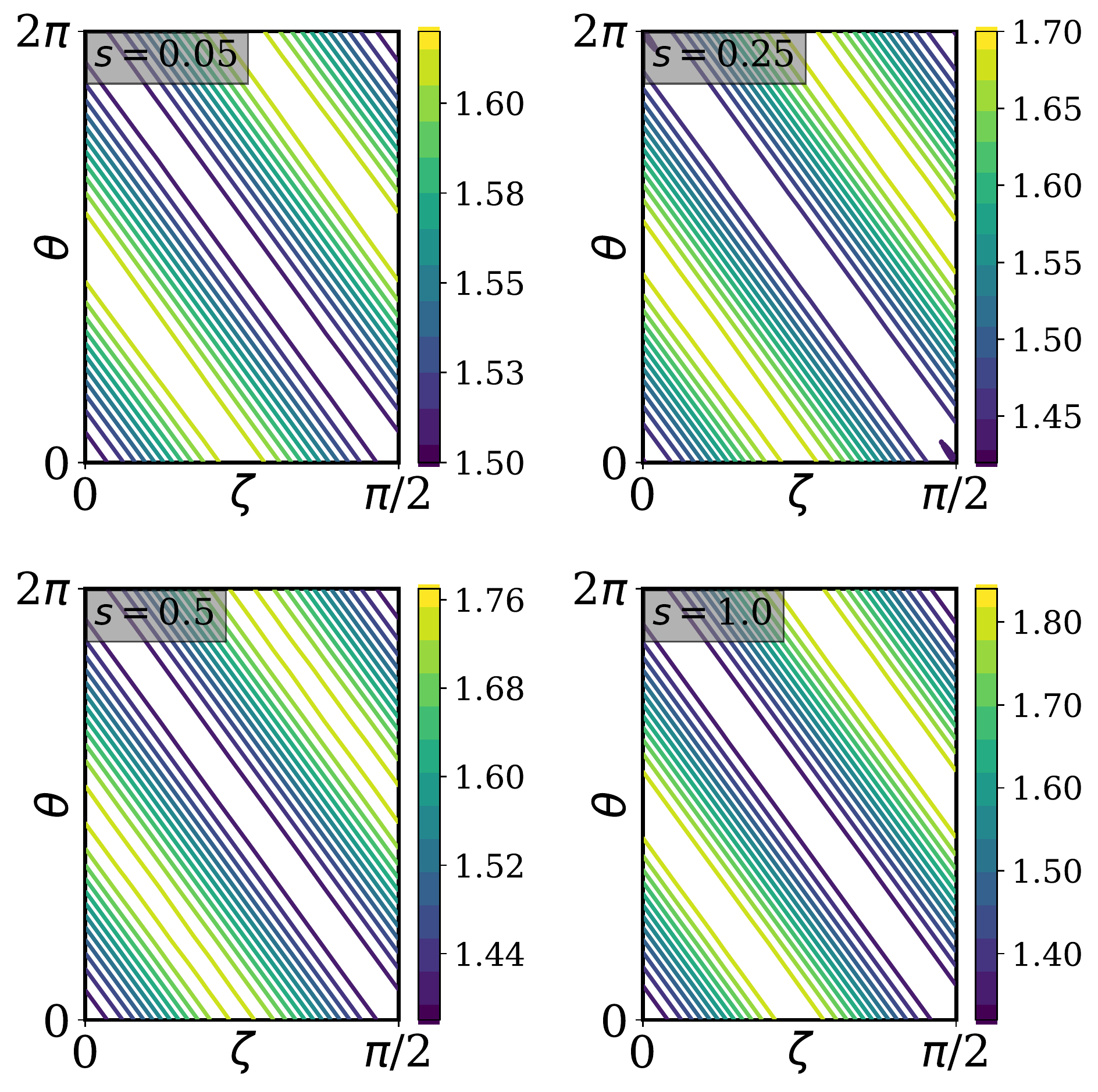}
\includegraphics[scale=0.3]{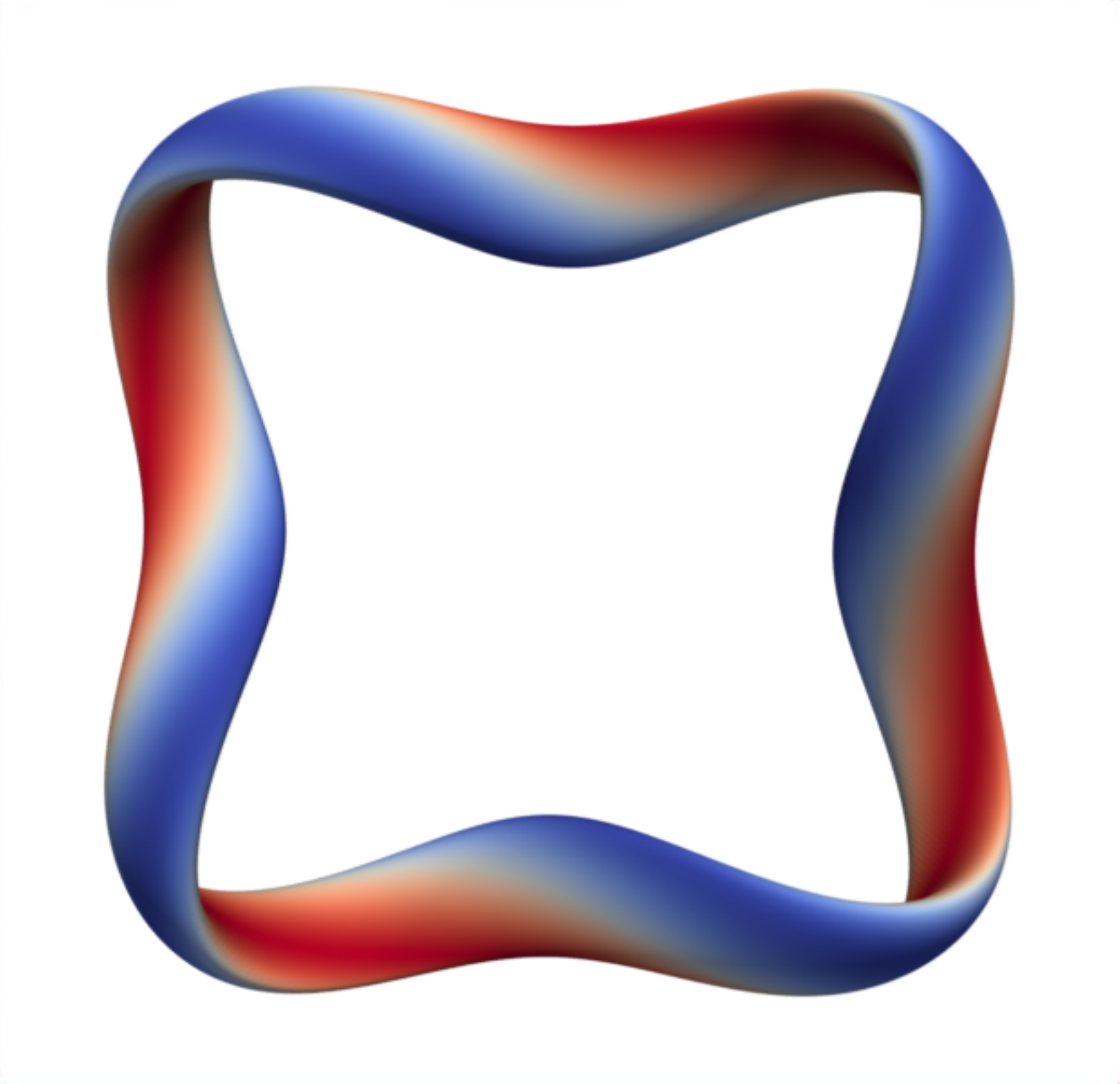}
\centering
\caption{(Left column) Contour plots of the magnetic field strength in Boozer coordinates $(\theta,\zeta)$ on four flux surfaces $s=0,0.25,0.5,1.0$ for three Pareto optimal configurations denoted by a star (aspect $3.5$), square (aspect $5.6$) and diamond (aspect $8.8$) in \Cref{fig:aspect_qs_pareto_front}. (Right column) Three dimensional renderings of corresponding Pareto optimal designs. The color of the 3D-configurations indicates the field strength, red is stronger.}
\label{fig:3d_plots}
\end{figure}

To reduce the computational overhead of using the $\epsilon$-constraint method we used the predictor-corrector method, introduced in \Cref{sec:continuation}, to explore the Pareto front between solutions of the $\epsilon$-constraint method. For the high dimensional decision space of interest, the full Hessians $\nabla^2 Q_{M,N}$ and $\nabla^2 A$ are too expensive to compute with finite differences. Instead we approximated the Hessian $\nabla^2 Q_{M,N}$ with a Gauss-Newton Hessian approximation and use a diagonal second-order central difference approximation to $\nabla^2 A$.

\Cref{fig:aspect_qs_pareto_front} shows the Pareto front for the problem \cref{eq:aspect_qs_prob}. \Cref{fig:3d_plots} shows three dimensional renderings of the three solutions highlighted by the square, star, and diamond markers in \Cref{fig:aspect_qs_pareto_front}, as well as contour plots of the field strength in Boozer coordinates. From \Cref{fig:aspect_qs_pareto_front} it is clear that low values of the quasi-symmetry violation, $Q_{1,-4}(\xb) < 10^{-3}$, can be achieved at all aspect ratios in our range. There is a slight trend indicating that quasi-symmetry may be achieved more precisely at higher aspect ratios. Nonetheless, by viewing the contour plots of the magnetic field strength in Boozer coordinates in \Cref{fig:3d_plots}, we see that configurations from all parts of the Pareto front have visibly precise quasi-symmetry.

While it seems that precise quasi-symmetry can be achieved at all aspect ratios considered, it is not clear how well the reduction in quasi-symmetry translates to an improvement in particle confinement. After all, achieving quasi-symmetry is a proxy for the true goal of achieving good confinement. To this end we computed the fraction of alpha particles lost from $10$ distinct Pareto optimal configurations, with aspect ratios 3, 3.5, 4, 4.5, 4.9, 5.6, 6.0, 8.5, 8.7, 9. To compute the losses we scaled the Pareto optimal configurations to the ARIES-CS reactor \citep{najmabadi2008aries} scale ($1.7$m minor radius and $B_{0,0}(s=0) = 5.7$T field strength on axis), and traced $5000$ particles born on the $s=1/4$ flux surface as well $5000$ particles born on the $s=1/2$ flux surface until a terminal time of $0.1$ seconds. Particles were traced according the vacuum guiding center equations in Boozer coordinates and were deemed lost if they crossed the $s=1$ flux surface. The loss fractions are shown in \cref{table:loss_fractions}. Not a single particle born on the $s=1/4$ flux surface was lost from nine of the ten configurations, and the remaining configuration, with aspect ratio $5.6$, lost only $0.22\%$ of the alpha particles. A slightly larger fraction of the alpha particles born on the $s=1/2$ flux surface were lost, between $0\%-2.6\%$ for each configuration. The loss fraction of alpha particles born on the $s=1/2$ flux surface increases slightly as the aspect ratio increases up until $A\approx 6$, then drops to approximately zero on the right tail of the Pareto front. It seems that there is no clear trend between the aspect ratio of the Pareto optimal designs and the loss fractions for particles born on either of the two surfaces. Thus while there is a slight trade-off between quasi-symmetry and aspect ratio, the trade-off between confinement and aspect ratio may be more complex. 

\begin{table}
  \begin{center}
  \def\arraystretch{1.2}
  \begin{tabular}{ll|rrrrrrrrrr}
    & & \multicolumn{10}{c}{Aspect Ratio}\\
    & & 3.0 & 3.5 & 4.0 & 4.5 & 4.9 & 5.6 & 6.0 & 8.5 & 8.7 & 9.0 \\
    \noalign{\hrule height 0.5pt} 
    \multirow{2}{*}{Surface $s$} & $1/4$ & 0.0 & 0.0 & 0.0 & 0.0 & 0.0 & 0.0022 & 0.0 & 0.0 & 0.0 & 0.0\\
     & $1/2$ & 0.010 & 0.015 & 0.018 & 0.021 & 0.023 & 0.026 & 0.026 & 0.0002 & 0.0002 & 0.0 
  \end{tabular}
  \caption{Fraction of alpha particles lost for ten Pareto optimal configurations from \Cref{fig:aspect_qs_pareto_front}. The loss fractions were computed by sampling $5000$ particles on the $s=1/4$ flux surface or $s=1/2$ flux surface and evolving their trajectories by the vacuum guiding center equations in Boozer coordinates until a terminal time of $0.1$ seconds or the particle breached the $s=1$ surface and was lost.}
  \label{table:loss_fractions}
  \end{center}
\end{table}

Spatially, we may expect or hope that the set of desirable configurations forms a compact region. However, we find the contrary -- the efficient set is not connected. Qualitatively, the large gap in the Pareto front in \Cref{fig:aspect_qs_pareto_front} hints that the efficient set is undergoing a \say{branch change}. Quantitatively, we measure the branch change by evaluating the Lagrange multipliers at edges of the gap in the Pareto front (roughly $A = 6.11, 8.5$) in \Cref{fig:aspect_qs_pareto_front}; a branch change is indicated by $\lambda = 0$, since this implies that the aspect ratio can be changed without causing any change in the quasi-symmetry. Indeed we find that the Lagrange multipliers are approximately zero, and hence that the efficient set corresponding to the small piece of the Pareto front, with aspect ratio greater than or equal to $8.5$, is disconnected from the remainder of the efficient set. While this may be a product of local optimization or the representation of decision variables, it is interesting to see nonetheless that the set of desirable configurations is disparate.

All of the Pareto optimal configurations found achieve exceptional levels of quasi-symmetry and particle confinement. From this it is clear that the aspect ratio is not limiting the device performance. On the other hand, engineering criteria on coil shape, fabrication tolerances, and placement tolerances may exhibit a greater trade-off with particle confinement, since the magnetic field generated by coils may not align well with the target magnetic field computed by Ideal MHD. In the next section we consider the trade-off between coil length, and the ability for coils to reproduce a target magnetic field.

\subsection{Problem 2: The coil length and quadratic flux trade-off}
Given the shape of the last close flux surface, $\Sc$, the \say{stage-two} stellarator optimization problem seeks to find magnetic coils with magnetic field $\Bb$, such that the field is orthogonal to the surface normal, $\Bb\cdot \nb=0$ \citep{merkel1987solution,zhu2017new}. Coils with a tight constraint on their length, will not be able to form intricate shapes and reduce the normal component of the magnetic field to zero, whereas coils which are longer can form arbitrary shapes that better minimize $\Bb\cdot \nb$. Long coils, on the other hand, are undesirable from a financial and engineering standpoint: longer coils are more expensive to build, often have higher curvature, and are more difficult to fit into the space around the device. In this experiment, we seek to understand the trade-off between coil length and the ability of coils to reproduce a target magnetic field. 

\begin{figure}
\includegraphics[scale=0.3]{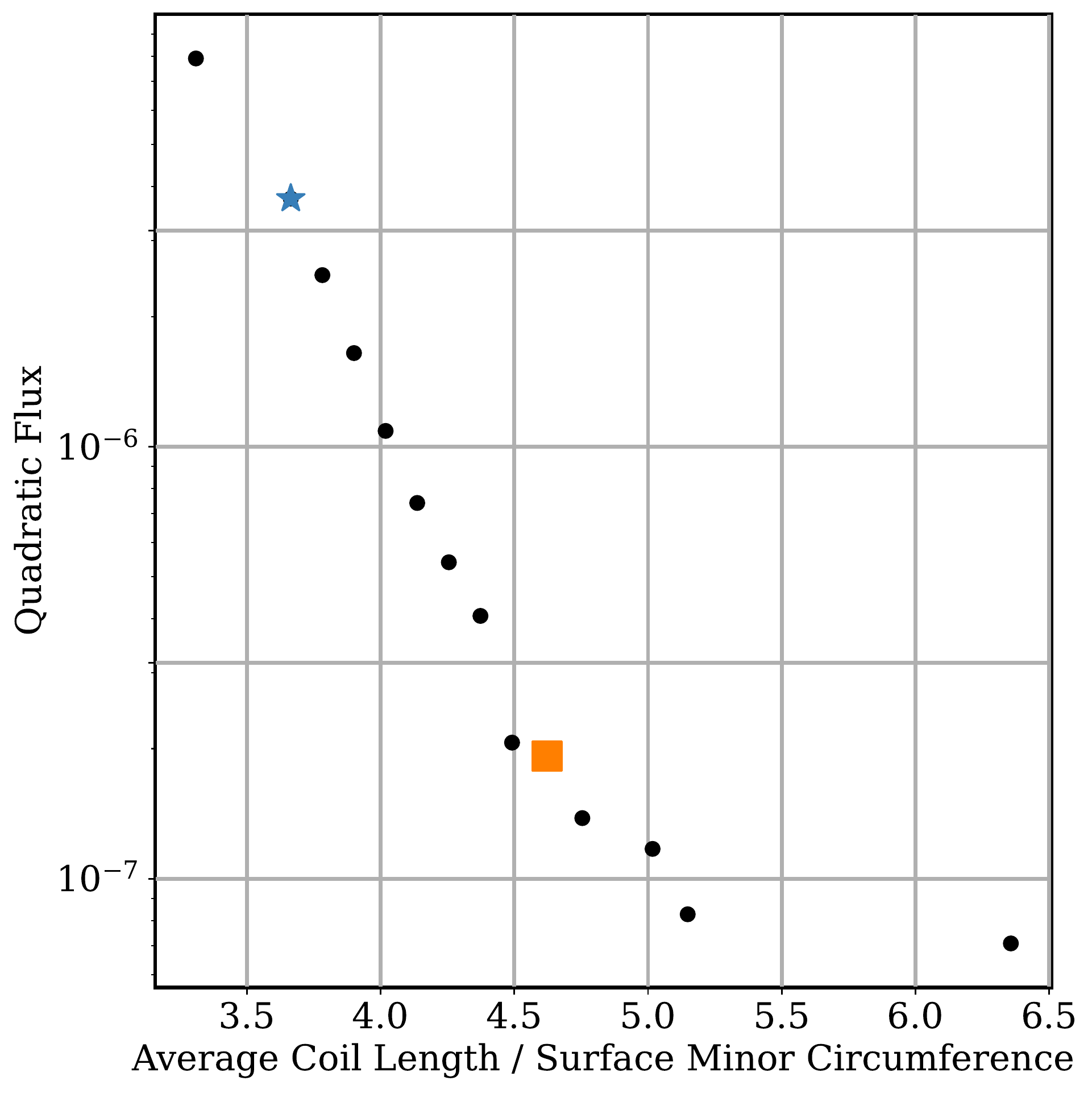}
\includegraphics[scale=0.25]{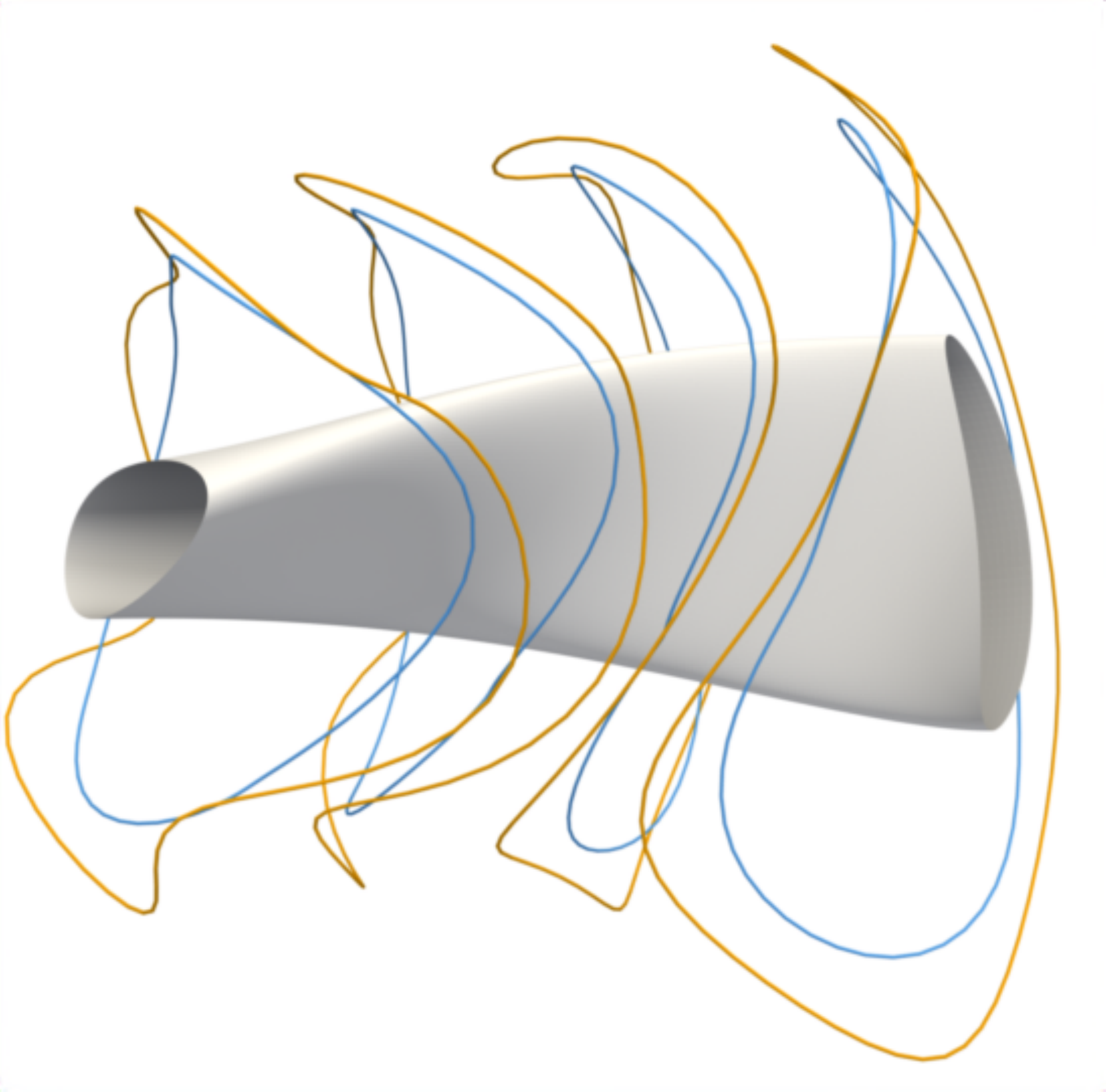}
\centering
\caption{(Left) Pareto front for coil length vs quadratic flux. The $x$-axis is non-dimensionalized by dividing by the minor circumference of the surface ($2\pi$ times the minor radius). (Right) Three dimensional rendering of the Pareto optimal coil sets denoted by the blue star (smaller blue coils) and the orange square (larger orange coils). }
\label{fig:coil_biobjective}
\end{figure}

We develop insight into this trade-off by solving the multi-objective problem
\begin{equation}
    \min_{\xb\in\Omegab} (J_{\nb}, L),
    \label{eq:coil_biobjective}
\end{equation}
where $J_{\nb}$ is the quadratic flux,
\begin{equation}
    J_{\nb}(\xb) = \frac{1}{2}\int_{\Sc}(\Bb\cdot \nb)^2 ds,
\end{equation}
$L$ is the total length of the $\nc$ coils considered,
\begin{equation}
    L = \sum_{i=1}^\nc L_i,
\end{equation}
and the decision variables $\xb$ represent coil shape parameters and the coil currents. 
The quadratic flux is a standard metric used to find coils in the stage-two problem \citep{merkel1987solution, zhu2017new,landreman2017improved,singh2020optimization,kruger2021constrained, glas2022global,wechsung2022stochastic}. To solve the bi-objective coil problem, \cref{eq:coil_biobjective}, we use the coil optimization framework implemented in \code{SIMSOPT}. The modular coils are described by a Fourier representation of the Cartesian coordinates with $\ncm$ modes. The Fourier representation of the $x$-coordinate of the $ith$ coil is,
\begin{equation}
    x^i(t,\xb) = x_{c,0}^i + \sum_{k=1}^{\ncm}[x_{c,k}^i\cos(kt) + x_{s,k}^i\sin(kt)],
\end{equation}
with analogous forms for the $y$ and $z$ coordinates. Each coil is additionally equipped with a current, though the current of the first coil is held fixed at $I_1 = 10^{5}$ Amperes throughout the optimization. The number of Fourier amplitudes per coil is $3(2\ncm + 1)$ making the total number of design variables $\nx = \nc3(2\ncm + 1) + \nc -1$. The coils considered satisfy field-period symmetry, as well as stellarator symmetry. Thus, the quadratic flux is only computed over a half-field-period, the total coil length is only taken over the coils in a half-field-period, and the decision variables $\xb$ only represent the $\nc$ coils in a half-field-period. The plasma boundary shape $\Sc$ used in this experiment is that of the $\nfp=2$ field-period quasi-axisymmetric(QA) configuration from \cite{landreman2022magnetic}, which we will refer to as LP-QA. $\ncm=5$ Fourier modes were used to describe the coil shapes.

The bi-objective problem was solved by applying the $\epsilonb$-constraint method, where quadratic flux was the objective and the total coil length was the constraint. The $\epsilonb$-constraint problems were solved with a quadratic penalty method \citep{nocedal1999numerical}. The penalty parameter was increased from an initial value of $1$ by a factor of $10$ at each iteration, and the quadratic penalty subproblems were solved by the L-BFGS-B algorithm \citep{zhu1997algorithm} with $2000$ iterations or until the norm of the gradient reached $10^{-6}$. 

\Cref{fig:coil_biobjective} shows the Pareto front for the bi-objective coil optimization problem \cref{eq:coil_biobjective} when $\nc=4$ coils are used per half-field-period. \Cref{fig:coil_biobjective} also shows three dimensional renderings of the two Pareto optimal coils sets, highlighted as the star and diamond in the plot of the Pareto front. \Cref{fig:coil_biobjective} shows that increasing the coil length substantially improves the ability for coils to reduce the quadratic flux up until the average coil length over minor circumference of the plasma boundary is about $5.5$, after which we numerically find no improvement. In addition we find that when the average coil length over minor circumference of the plasma boundary exceeds roughly $5$, the coils become exceedingly complex: coils curvature becomes large and coils begin to pass under one another, competing for space. At these coil lengths, coil curvature and the coil-to-coil separation should also be constrained.

\begin{figure}
\includegraphics[scale=0.2]{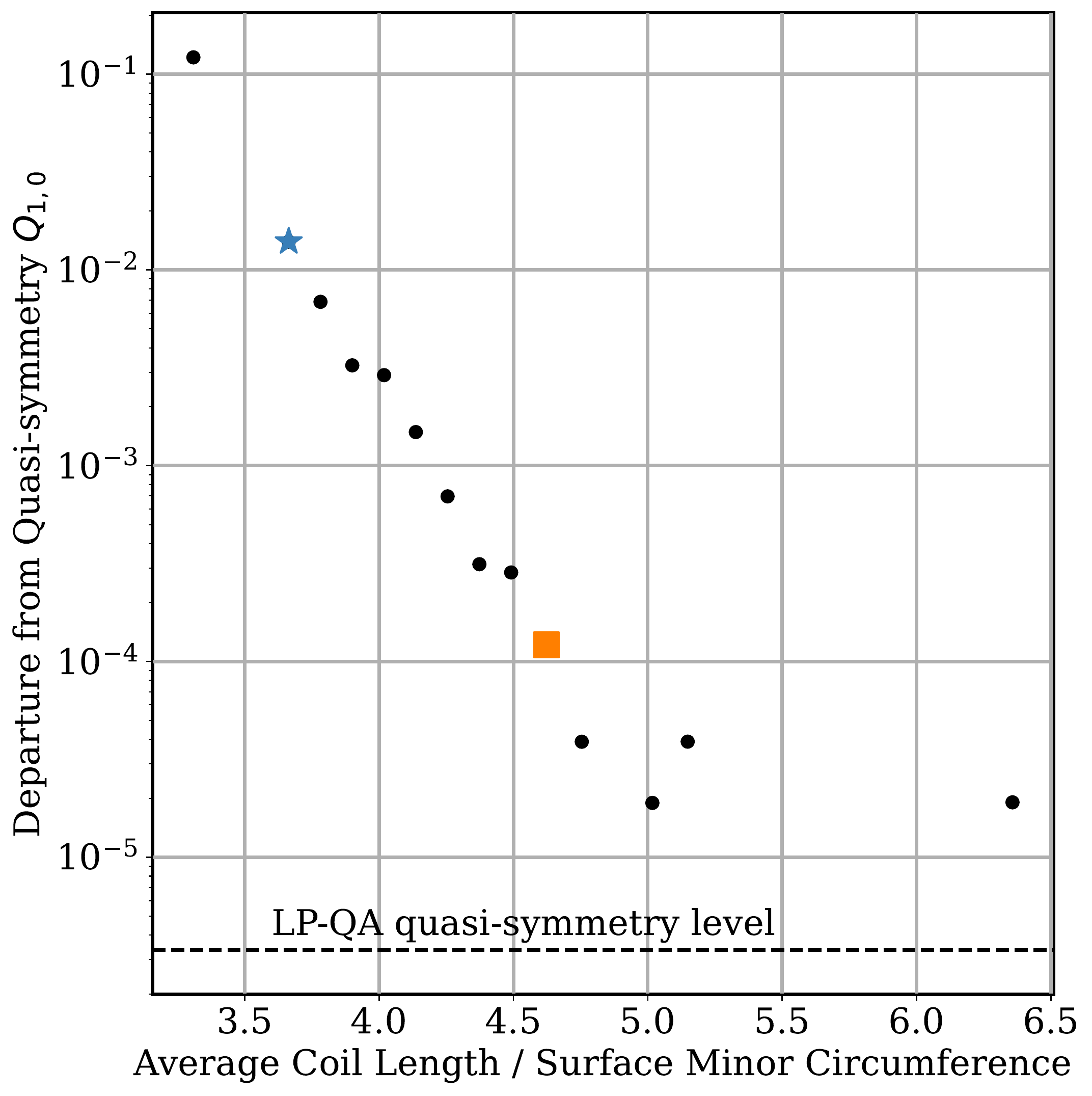}
\includegraphics[scale=0.45]{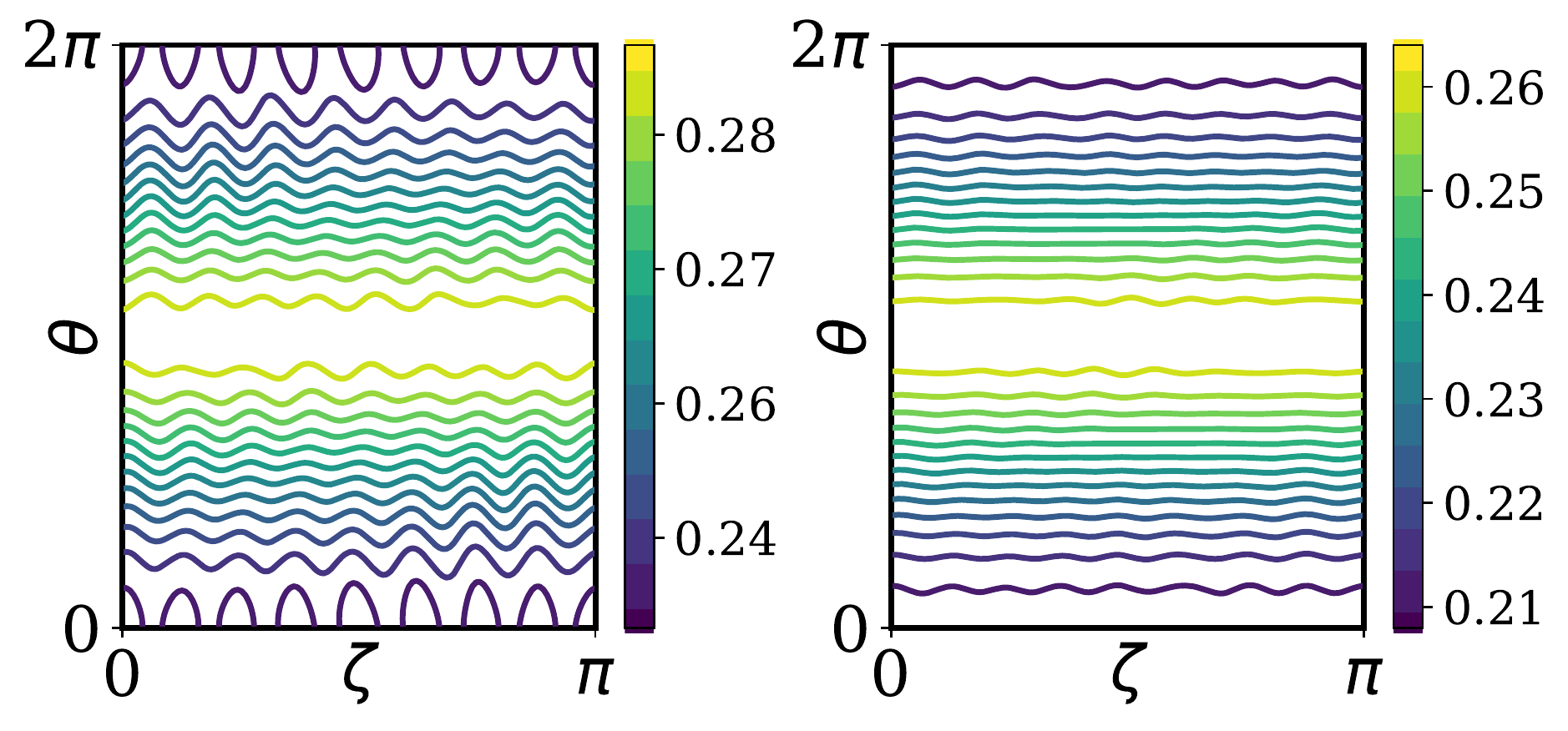}
\centering
\caption{(Left) Trade-off between quasi-axisymmetry and coil length for the Pareto optimal points shown in \Cref{fig:coil_biobjective}. For each Pareto optimal coil set, the quasi-axisymmetry metric $Q_{1,0}$ was computed by running \code{VMEC} with a boundary surfaced obtained by fitting a quadratic flux minimizing surface to the field generated by the coils. (Middle and right) Contour plots of the coil generated field strength in the Boozer toroidal and poloidal angles $\zeta,\theta$ on the $s=1$ surface of the LP-QA configuration for the coil sets of the blue star and orange square shown in \Cref{fig:coil_biobjective}.}
\label{fig:qfm_plot}
\end{figure}

While \Cref{fig:coil_biobjective} shows that constraining the total coil length limits the ability for coils to reproduce a magnetic field, it does not show how the limits on the total coil length relate to a loss in quasi-symmetry or loss in confinement. In \Cref{fig:qfm_plot} we measure the extent to which the Pareto optimal coil configurations from \Cref{fig:coil_biobjective} generate quasi-symmetric magnetic fields. To do so, we fit a quadratic flux minimizing (QFM) surface \citep{dewar1994almost} to the coil-generated magnetic field $\Bb$, evaluate \code{VMEC} using the QFM surface as the boundary shape, and compute the quasi-symmetric metric $Q_{1,0}$. The QFM surface is constrained to have identical volume to the LP-QA target surface. We find the QFM surface by solving the following problem in \code{SIMSOPT}:
\begin{align}
    &\min_{S} \frac{\int_S(\Bb\cdot \nb)^2 ds}{\int_S\|\Bb\|^2 ds},
    \\
    \text{s.t. } &\text{Volume}(S) = \text{Volume}(\text{LP-QA}).
\end{align}
\Cref{fig:qfm_plot} shows that the quasi-axisymmetry degrades significantly, by three orders of magnitude,  as the coil length constraints are tightened. Unlike the longer coil sets, the shorter coil sets generate significant coil ripple, degrading the quasi-symmetry. While it may not be possible to entirely avoid the loss of quasi-symmetry when coil constraints are tightened, the loss may be diminished by using single-stage optimization approaches to design the coils. Part of the loss of quasi-symmetry is inherent to the two-stage approach to coil design: the coils are optimized to reproduce a target magnetic field, but not to generate a quasi-symmetric field. Single-stage optimization approaches \citep{giuliani2022single,giuliani2022direct,jorge2023single} on the other hand, directly optimize coils for quasi-symmetry, ensuring that they achieve optimal quasi-symmetry levels over the space of coils satisfying the coil length constraint. Single stage approaches would hence improve the unfortunate trade-off between coil length and quasi-symmetry.

\section{Discussion}
\label{sec:discussion}

Understanding trade-offs in stellarator designs is particularly important in designing high performance devices that satisfy the multitude of physical, engineering, and financial criteria. Throughout this study we have shown how multi-objective optimization can be used to investigate trade-offs and develop insight into the role of design parameters.

The example problems considered here were biobjective problems, but many stellarator design problems may have three or more competing objectives. MOO methods are useful in this context, however the dimension of the Pareto front grows with the number of objectives which makes visualizing the Pareto front difficult and thoroughly exploring the Pareto front often intractable. For this reason, we generally recommend exploring trade-offs between two or three objectives at a time. For three objectives, scalarization methods \citep{ehrgott2005book,miettinen2012book} and other gradient-based MOO methods \citep{fliege2000steepest,desideri2012multiple,schaffler2002stochastic,peitz2018gradient,gkaragkounis2018adjoint} are an appropriate choice since they allow the use of derivatives and for the user to dicate which part of the Pareto front is explored. For problems with more than three objectives, parallel algorithmic approaches will find promising configurations most efficiently \citep{chang2023parmoo, knowles2006parego, daulton2022multi, wagner2010expected, deb2002fast}. 

Looking beyond the two trade-offs considered here, there are a host of other trade-offs that appear in stellarator design that deserve attention. There is a natural trade-off, for instance, between coil complexity and quasi-symmetry that should be explored using a direct single state optimization method, such as with the near-axis expansion \citep{wechsung2022single}. In addition, it is not well understood how stability criteria trade off with particle confinement criteria, or how design flexibility for multi-purpose coils trades off with volume and quasi-symmetry \citep{lee2022stellarator}. Ideally, these problems should be solved and the Pareto optimal solutions should be tabulated in a way that practitioners can easily survey a multitude of configurations and analyze their strengths and weaknesses as a holistic set.

\section{Data Availability}
Code and data can be found at \url{https://doi.org/10.5281/zenodo.7838063}.

\section{Acknowledgements}
This work was generously supported 
by a grant from the Simons Foundation (No. 560651, D.B.).

\bibliographystyle{jpp}
\bibliography{references}

\end{document}